\documentclass[aps,prl,preprint,a4paper]{revtex4}
\usepackage{amsmath}
\usepackage{amsfonts}
\usepackage{graphicx}

\begin{document}

\preprint{ITP-UH-11/05}

\title{Correlation functions of one-dimensional Bose-Fermi mixtures}

\author{Holger Frahm}
\author{Guillaume Palacios}
\affiliation{Institut f\"ur Theoretische Physik, Universit\"at Hannover,
             Appelstr.\ 2, 30167 Hannover, Germany}

\date{July 14, 2005}

\begin{abstract}
We calculate the asymptotic behaviour of correlation functions as a function
of the microscopic parameters for a Bose-Fermi mixture with repulsive
interaction in one dimension.  For two cases, namely polarized and unpolarized
fermions the singularities of the momentum distribution functions are
characterized as a function of the coupling constant and the relative density
of bosons.
\end{abstract}
\maketitle

In the past years, the advances in cooling and trapping of atomic gases have
opened the possibility to realize quasi one-dimensional (1D) systems with
tunable strength of the interactions in optical lattices.  This gives rise to
new opportunities for the investigation of the striking phenomena appearing in
correlated systems as a consequence of the enhanced quantum fluctuations in
reduced spatial dimensions.  The observable signatures of these phenomena are
encoded in the correlation functions of the system such as the momentum
distribution function which can be measured directly in time-of flight
experiments or using Bragg spectroscopy \cite{EXPnk}.  A setup for the
measurement of various density correlation functions has been proposed for the
identification of dominant correlations in the atomic gas
\cite{CORR}.  Theoretical studies of correlation functions in cold
atomic gases have been performed both using analytical methods, e.g.\
bosonization \cite{GoNeTs:boson} combined with exact results from integrable
models such as the Bose-gas with repulsive $\delta$-function interaction
\cite{BOSE1d}, and numerically.
New correlation effects appear when the particles considered have internal
degrees of freedom.  In cold gases containing different constituent atoms
Bose-Fermi mixtures can be realized \cite{MIX}.  Extensive theoretical results
exist for 1D Fermi gases due to their equivalence with the Tomonaga-Luttinger
(TL) liquids realized by correlated electrons in 1D lattices
\cite{GoNeTs:boson,TheBook}.  Only recently, theoretical investigations have
been extended to Bose-Fermi mixtures: some correlation functions have been
calculated numerically in the strong coupling limit \cite{ChSZ05,ImDe05} where
the problem simplifies due to the factorization of the many-particle wave
function (see e.g.\ \cite{OgSh90}).
For analytical results on these systems one has to go beyond mean-field
approximations and use methods which can capture the strong quantum
fluctuations in 1D systems.  The phase diagram and certain correlation
functions of atomic mixtures have been studied in the Luttinger liquid picture
\cite{CaHo03,MatX04}.  Without further input, however, these results are
limited to the weakly interacting regime since the TL parameters which
determine the low-energy theory cannot easily be related to the microscopic
parameters describing the underlying gas.  Therefore, instabilities predicted
within this approach may not appear in a specific realizations
\cite{LaYa71,ImDe05,BatX05b}.

In this paper we establish the relation between the TL and the microscopic
parameters for an integrable Bose-Fermi mixture \cite{LaYa71}.  We employ
methods from conformal quantum field theory (CFT) to determine the asymptotic
(long distance, low-energy) behaviour of correlation functions in the model
from a finite size scaling analysis of the exact spectrum obtained by means of
the Bethe ansatz.  This approach gives the complete set of critical exponents
of the model as a function of the parameters in the microscopic Hamiltonian
(see e.g. \cite{FrKo,KaYa91,TheBook} for applications to 1D
correlated electrons).  As an application we compute the momentum distribution
function of bosons and fermions in the atomic mixture as a function of their
respective densities and the effective coupling constant.
It should be emphasized that our results can be expected to describe the
generic (universal) low-energy behaviour of atomic mixtures.  Additional
interactions -- as long as they do not lead to a phase transition -- will
merely change the anomalous exponents but not the qualitative behaviour of the
correlation functions.


The 1D Bose-Fermi mixture of $N=M_f+M_b$ particles with repulsive interaction
($c>0$) on a line of length of length $L$ subject to periodic boundary
conditions is described by the integrable Hamiltonian \cite{LaYa71}
\begin{equation}
  {\cal H} = -\sum_{i=1}^N \frac{\partial^2}{\partial x_i^2}
             + 2c\sum_{i<j} \delta(x_i-x_k)\ . 
\label{hamil}
\end{equation}
Here $M_f=M_\uparrow+M_\downarrow$ of the particles are fermions carrying spin
$\sigma=\uparrow,\downarrow$ and $M_b$ of them are bosons.  The many-particle
eigenstates of (\ref{hamil}) are parametrized by the solutions of the Bethe
ansatz equations (BAE) \cite{LaYa71}
\begin{align}
 & \exp({iq^{(0)}_j L}) = \prod_{k=1}^{M_1} e_c(q^{(0)}_j-q^{(1)}_k)
\notag\\
 & \prod_{j=1}^{M_0} e_c(q^{(1)}_k-q^{(0)}_j)
  \prod_{\ell=1}^{M_2} e_c(q^{(1)}_k-q^{(2)}_\ell) =
  \prod_{k'\ne k}^{M_1} e_{2c}(q^{(1)}_{k}-q^{(1)}_{k'})
\notag\\
 & \prod_{k=1}^{M_1} e_c(q^{(2)}_\ell-q^{(1)}_k) = 1
\label{bae}
\end{align}
where $e_a(x) = (x+ia/2)/(x-ia/2)$ and $M_0=N$, $M_1=N-M_\uparrow$,
$M_2=M_b$.  The corresponding eigenvalue of (\ref{hamil}) is $E=\sum_{j=1}^N
(q_j^{(0)})^2$.
In the thermodynamic limit $L\to\infty$ with $M_i/L$ kept fixed, the root
configurations $\{q^{(i)}\}$ of Eqs.~(\ref{bae}) can be described by
distribution functions $\rho_i$ which, as a consequence of (\ref{bae}) are
solutions to \cite{LaYa71}
\begin{equation}
  \rho_i(x) = \rho^{(0)}_i + \sum_j (\hat{K}_{ij}\otimes\rho_j)(x)
\label{irho}
\end{equation}
Here $\rho^{(0)}_i = (c/2\pi) \delta_{i0}$ and $\hat{K}_{ij}$ are linear
integral operators acting as ($\int_j \equiv \int_{-Q_j}^{Q_j} {\mathrm{d}}y$)
\begin{equation}
  (\hat{K}_{ij} \otimes f)(x) = 
  \int_j k_{ij}(x-y) f(y)\ .
\end{equation}
The kernels of these integral operators are $k_{ij}(x) = a_1(x)
\delta_{|i-j|,1} - a_2(x) \delta_{i,1}\delta_{j,1}$
where $2\pi a_n(x) = 4n/(4x^2+n^2)$.  The properties of the system are
completely characterized by the densities $m_i\equiv M_i/L = \int_i
\rho^{(i)}(x)$ of the components of the mixture (these relations determine the
boundaries $Q_i$ of the above integral equations) and the dimensionless
coupling strength $\gamma = Lc/N$.  For later use we also introduce the
fraction of bosons in the system $\alpha=M_b/N$.


Generically, i.e.\ for $m_i>0$, there are three modes of collective elementary
excitations above the many-particle ground state of (\ref{hamil}).  
Their dispersion $\epsilon_i(k)$ 
is linear at low energies with different sound velocities $v_i=\partial
\epsilon_i/\partial k$, $i=0,1,2$ \cite{BatX05b}.  These quantities determine
the finite size scaling behaviour of the ground state energy
\begin{equation}
  E_0 - L\epsilon_\infty = -\frac{\pi}{6L} \sum_i v_i +
  \mathit{o}\left(\frac{1}{L}\right)\ . 
\label{fs0}
\end{equation}
Physical excitations of the system are combinations of the elementary ones.
Due to the interacting nature of the system the different modes are coupled
and excitations in one of the modes shift the energies in the other ones.  In
general, this effect can be described in terms of generalized susceptibilities
which may be determined in an experiment or numerically from studies of small
systems \cite{TheBook}.  For the Bethe ansatz solvable models it is possible
to describe the coupling of the modes in terms of the dressed charge matrix
\cite{DCREF} which in this case reads
\begin{equation}
  Z_{ij} = \xi_{ij}(Q_i)\ .
\label{zmatrix}
\end{equation}
The functions $\xi_{ij}$ are given in terms of integral equations 
  $\xi_{ij}(x) = \delta_{ij} + \sum_k (\hat{K}_{ik}\otimes\xi_{kj})(x)$.

$Z$ determines the general form of the finite size corrections to the energies
of low-lying excitations
\begin{align}
  &\Delta E(\Delta\mathbf{M},\mathbf{D}) = {2\pi\over L}\Bigl(
  {1\over4} \Delta \mathbf{M}^\top
	 \left(Z^\top\right)^{-1} V Z^{-1} \Delta \mathbf{M}
\notag\\
  & + \mathbf{D}^\top Z V Z^{\top} \mathbf{D}
    + \sum_{k} v_k \left(N_k^+ + N_k^-\right)\Bigr)
    +\mathit{o}\left(1\over L\right)\ .
\label{fse}
\end{align}
Here, $V=\mathrm{diag}(v_0,v_1,v_2)$ is a $3\times3$ matrix of the sound
velocities, $N_k^\pm$ are non-negative integers, $\Delta M$ is a vector of
integers denoting the change of $M_i$ with respect to the ground state for
charged excitations.  The $D_i$ are integers or half-odd integers according to
\begin{align}
  D_0 \sim \left( \Delta M_0+\Delta M_1\right)/2 
      = \Delta M_\uparrow/2 \mod 1
\notag\\
  D_1 \sim \left( \Delta M_0+\Delta M_2\right)/2 
      = \Delta M_f/2 \mod 1
\label{selrul}\\
  D_2 \sim \left( \Delta M_1+\Delta M_2\right)/2 
      = \Delta M_\downarrow/2 \mod 1
\notag
\end{align}
and enumerate finite momentum transfer processes:
\begin{multline}
 \Delta P(\Delta\mathbf{M},\mathbf{D}) = {2\pi\over L} \left(
   \Delta \mathbf{M}^\top \cdot {\mathbf D} + 
   \sum_{k}\left(N_k^+ - N_k^-\right)\right) \\
 \qquad + 2 {k}_{F,\uparrow} D_0 + 2 
   {k}_{F,\downarrow}\left( D_0 + D_1\right) + 2 k_B \sum_j D_j\ .
\label{fsp}
\end{multline}
Here $k_{F,\sigma} = \pi M_\sigma/L$ are the Fermi momenta of the fermion
components, $k_B = \pi M_b/L$ is the corresponding quantity for the
interacting bosons.

In the framework of CFT \cite{BPZ} the finite size spectrum (\ref{fs0}),
(\ref{fse}) can be understood as that of a critical theory based on the
product of three Virasoro algebras each having central charge $C=1$
\cite{FrKo,TheBook}.  Correlation functions of a general operator in the
theory -- characterized by the quantum numbers $\Delta M_i$ and $D_i$ -- will
contain contributions from these three sectors.  The simplest ones, analogues
of primary fields in the CFT, have correlation functions (in Euclidean time
$\tau$)
\begin{widetext}
\begin{equation}
  \left\langle \phi_{\Delta}(x,\tau) \phi_{\Delta}(0,0) \right\rangle =
  \frac{\exp\left(2i D_0 {k}_{F,\uparrow}x
  +2i \left(D_0+D_1\right)  {k}_{F,\downarrow}x 
  +2i \left(\sum_j D_j\right) k_B x\right)
  }{\prod_{k} (v_k\tau +ix)^{2\Delta^+_k}(v_k\tau -ix)^{2\Delta^-_k}}\ .
\label{corr0}
\end{equation}
\end{widetext}
The operators $\phi_{\Delta}$ are characterized by their scaling dimensions
$\Delta_k^\pm$ in the chiral (left- and right moving) components of all three
constituent theories.  The latter are uniquely determined from the finite size
energies (\ref{fse}) and momenta (\ref{fsp}) and form towers starting at
\begin{align}
  2\Delta_k^\pm = \left( \sum_j Z_{kj} D_j \pm
       \frac{1}{2}\sum_j \Delta M_j (Z^{-1})_{jk} \right)^2 .
\label{cdim}
\end{align}
The asymptotic exponential decay of correlation functions in a large but
finite system or at finite temperature $T$ can be obtained from (\ref{corr0})
by conformal invariance.  For example, at $T>0$ the denominators in
(\ref{corr0}) have to be replaced by by
$
  (v_k\tau \pm ix)^{-2\Delta^\pm_k} \to 
      \left({\pi T}/v_k{\sin \pi T (\tau \pm ix/v_k)}\right)^{2\Delta^\pm_k}
$.

With (\ref{cdim}) the critical exponents which determine the long-distance
asymptotics of \emph{any} correlation function are known as soon as we have
computed the dressed charge matrix (\ref{zmatrix}).
To calculate the correlation functions of a given local operator ${\cal O}$ in
the microscopic theory (\ref{hamil}) one needs to know its expansion in terms
of the fields $\phi_{\Delta}$ of the CFT.  Usually, this expansion is not
known but ${\cal O}$ and $\phi_{\Delta}$ have to generate the same set of
selection rules in calculating the correlation function.  This drastically
reduces the number of possible terms in the expansion: As an example consider
the bosonic Green's function $G_b(x,\tau)= \langle \Psi_b(x,\tau)
\Psi_b^\dagger(0,0)\rangle$: clearly $\Psi_b^\dagger$ generates a state with
$\Delta M_b=1$ which implies $\Delta M_j\equiv1$ in (\ref{cdim}).  By
(\ref{selrul}) the quantum numbers $D_j$ are further restricted to integers:
the uniform part of the $G_b(x) \sim |x|^{-1/2K_b}$ is described by the
operator with $D_j\equiv0$ which allows to identify the TL parameter $K_b$
\cite{HALD} from (\ref{cdim}).  The interactions lead to additional
contributions to $G_b$ oscillating with wave numbers $k_0=2k_{F\sigma}$,
$2k_B$, \ldots

For a comparison with experimental data one is often interested in Fourier
transforms of the two-point correlation functions given above.  The large
distance behaviour of (\ref{corr0}) determines the singularities of spectral
functions near $\omega\approx \pm v_k(k-k_0)$ (see e.g.\ \cite{TheBook}).
Quantities accessible in experiments with cold gases \cite{EXPnk} are
the momentum distribution functions of the constituent particles.  For the
bosons this is the Fourier transform of the equal time Green's function
$G_b(x)$.  From (\ref{corr0}) its singularities at wave numbers $k_0$ are then
$n_b(k) \sim |k-k_0|^{\nu_b}$ near $k\approx k_0$.  The exponent $\nu_b$ is
the minimal value of $2\left(\sum_k \Delta_k^++\Delta_k^- \right) -1$
compatible with the quantum numbers $\Delta M$ and the selection rules for the
$D$ for the given $k_0$, e.g.\ $1/2K_b=\nu_b+1=\frac{1}{4}
\sum_k(\sum_j (Z^{-1}_{jk}))^2$ for $k_0=0$.

Using the same procedure for the fermionic Green's function $G_\sigma(x,\tau)
= \langle\Psi_\sigma(x,\tau)\Psi_\sigma^\dagger(0,0)\rangle$ we find that
their asymptotic behaviour is determined by the conformal fields with $\Delta
M_0=1$, $\Delta M_1 = \Delta M_2 = 0$, half-odd integers $D_0$, $D_1$ and
integer $D_2$ for $G_\uparrow$ and $\Delta M_0=1=\Delta M_1$, $\Delta M_2 =
0$, half-odd integers $D_1$, $D_2$ and integer $D_0$ for $G_\downarrow$.
Again, the singularities of the fermions' distribution functions $n_\sigma(k)$
follow from (\ref{corr0}).  
Near $k_0-k_{F\sigma}=0$, $\pm 2k_B$, \ldots they are given by
\begin{equation}
  n_\sigma(k) \sim \mathrm{sign}(k-k_0)|k-k_0|^{\nu_f},\quad
    \mathrm{for~}k\approx k_0\ .
\end{equation}
$\nu_f$ is related to the dimensions (\ref{cdim}) for the quantum
numbers $\Delta M$ and $D$ as $\nu_b$ above.  The Fermi distribution of
non-interacting particles corresponds to $\nu_f(k_{F\sigma})=0$.


In the following we consider two cases of particular relevance
\cite{LaYa71,MatX04,BatX05b}, namely (i) the unpolarized case where
$M_\uparrow=M_\downarrow=M_f/2$ and the ground state of the system is
invariant under rotations in the spin index of the fermions and (ii) the fully
polarized case where there is only one spin component of the fermions.

\paragraph{The unpolarized gas.}
For $Q_1=\infty$ one obtains $M_\uparrow=M_\downarrow$ from (\ref{irho}),
i.e.\ with vanishing net magnetization.  In this case the dressed charge
matrix (\ref{zmatrix}) takes the form
\begin{equation}
  Z = \frac{1}{2}\left(\begin{array}{ccc}
         2\zeta_{00} & (\zeta_{00}+\zeta_{01}) & 2\zeta_{01} \\
              0    & {\sqrt{2}} & 0 \\
         2\zeta_{10} & (\zeta_{10}+\zeta_{11}) & 2\zeta_{11}
	    \end{array}\right)\ .
\label{zmat00}
\end{equation}
Here the Wiener-Hopf method has been used to determine $Z_{11}=1/\sqrt{2}$ and
$\zeta_{0j} = \zeta_{0j}(Q_0)$, $\zeta_{1j} = \zeta_{1j}(Q_2)$.  The functions
$\zeta_{ij}(x)$ are given by
\begin{equation}
  \zeta_{ij}(x) = \delta_{ij} + \int_0 R(x-y) \zeta_{0j}(y)
                              + \int_2 R(x-y) \zeta_{1j}(y)\
\label{dc_up}
\end{equation}
with $R(x) = (1/\pi) \int_0^\infty \mathrm{d}\omega\ \mathrm{e}^{-|\omega|/2}
\cos(\omega x)/\cosh(\omega/2)$.  Using (\ref{zmat00}) the scaling dimensions
$\Delta_1^\pm$ in (\ref{cdim})
are independent on the remaining system parameters, i.e.\ the effective
coupling $\gamma$ and the bosonic fraction $\alpha$.  This a consequence of
the $SU(2)$ invariance of the system in this case.  The mode $\epsilon_1(k)$
is the spinon mode of the unpolarized system, the CFT describing its
low-energy properties is an $[SU(2)]_1$ Wess-Zumino-Witten model.

\begin{figure}
\includegraphics[width=0.5\textwidth]{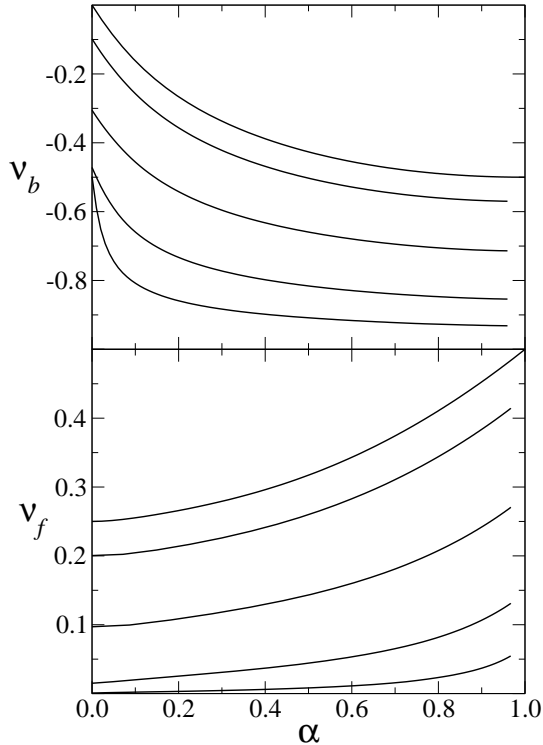}
\caption{
  \label{fig-up}
  Exponents characterizing the singularities in the bosonic (upper panel) and
  fermionic (lower panel) momentum distribution function for the unpolarized
  gas at $k=0$ and $k_F$, respectively, as a function of the bosonic fraction
  $\alpha$ in the mixture for $\gamma=0.2$, $1.0$, $5.0$, $25.0$, $\infty$
  (bottom to top).}
\end{figure}

Additional simplifications arise in the strong coupling limit
$\gamma\to\infty$ (i.e.\ $Q_0\to0$) where $\zeta_{00}=1$, $\zeta_{10}=0$,
$\zeta_{10}=\zeta_{11}(0)-1=\alpha$ and $\zeta_{11}(x)$ is given by a scalar
integral equation resulting from (\ref{dc_up}).
In Fig.~\ref{fig-up} we present results obtained from the numerical solution
of these integral equations for the exponents which determine the
singularities of the momentum distribution functions for bosons at $k=0$ and
fermions at $k=k_F$ as a function of the bosonic fraction $\alpha$ for various
values of $\gamma$.  The exponents $\nu_{b,f}$ at the other wave numbers are
always larger than 1.
Note that the system is in a different universality class for $\alpha=0$ or
$1$.  At $\alpha=0$, all particles are fermionic and the critical exponents
are those of the 1D Fermi gas \cite{TheBook}.  Here the exponent $\nu_f$ for
the singularity at the Fermi point varies between $0$ and $1/8$ as a function
of $\gamma$.  On the other hand, the limit of $\nu_b$ as $\alpha\to1$ gives
exactly the exponent of the 1D Bose gas with $\delta$-interaction \cite{HALD}.

\begin{figure}
\includegraphics[width=0.5\textwidth]{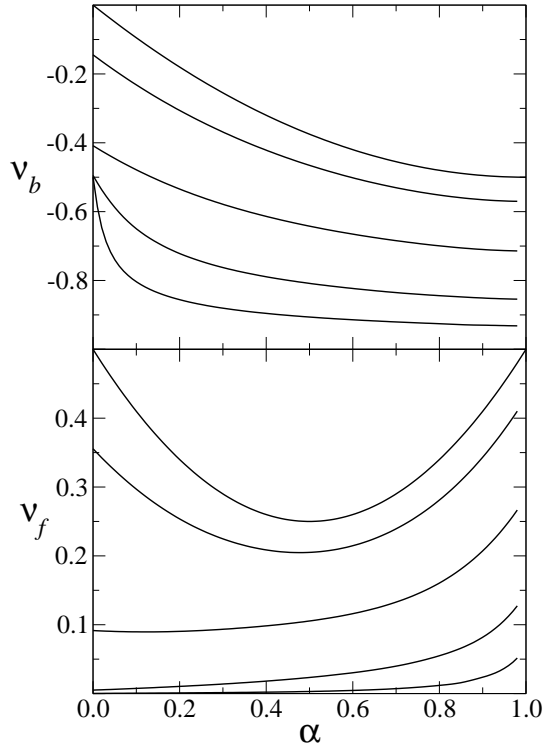}
\caption{
  \label{fig-pp}
  Same as Fig.~\ref{fig-up} for a mixture with polarized fermions.}
\end{figure}

\paragraph{The spin-polarized gas.}
Setting $Q_2=\infty$ in (\ref{irho}) corresponds to $M_\downarrow=0$.  This
case has been discussed recently in Ref.~\onlinecite{ImDe05} where some
correlation functions have been computed numerically in the strong coupling
limit.  
In this case, the finite size spectrum and the scaling dimensions are
determined by two gapless modes.  Again, the equations simplify in the strong
coupling limit where all exponents can be given as a function of $\alpha$
directly, e.g.\ $\nu_b(0)=\alpha^2/2-\alpha$ and
$\nu_f(k_F)=\alpha^2-\alpha+1/2$ for the dominant singularities of the bosonic
and fermionic momentum distribution functions, respectively (see
Fig.~\ref{fig-pp} for the $\gamma$ dependence).
While the dependence of $\nu_b$ on $\alpha$ is similar to the one found in the
unpolarized case, the strong coupling behaviour of $\nu_f$ at small bosonic
fraction is seen to be very different.  
Note that the singularity at $k_F+2k_B$ becomes very pronounced for
sufficiently small $\alpha$ ($\nu_f=\alpha^2+\alpha+1/2$ at strong coupling).
This feature of the fermionic distribution function is a direct signature of
the interaction and should be observable in experiments.

In summary we have used predictions from CFT on the finite size scaling of the
low-energy spectrum to compute the critical properties of a 1D Bose-Fermi
mixture for the integrable model (\ref{hamil}).  In the generic case there are
three linearly dispersing modes which determine the low-energy effective
theory and the asymptotic behaviour of the correlation functions.  Within this
formalism we have related the critical exponents directly to the parameters
describing the microscopic Hamiltonian, i.e.\ the coupling strength, the
fraction of bosons and polarization of the fermions.  
The approach can be used to investigate the phase diagram of the 1D mixture by
identifying the order parameter with the slowest long-distance decay of its
correlation functions (smallest exponent) \cite{CaHo03,MatX04}.  Within the
integrable model there is no instability leading to a phase transition
\cite{LaYa71,BatX05b}.  The general expression (\ref{cdim}), however, allows
to obtain estimates on the exponents in a more general system based on
numerical data on the spectrum of finite systems.

\begin{acknowledgments}
This work has been supported by the Deutsche Forschungsgemeinschaft.
\end{acknowledgments}


\end{document}